\documentstyle[12pt]{article}

\begin{document}

\def\e{\varepsilon}

\def\ss{s}         \def\sb{\bar s}
\def\qq{q}         \def\qb{\bar q}
\def\mud{\dot \mu} \def\nud{\dot \nu}
\def\ub{\bar u}    \def\vb{\bar v}

\def\|{\,}


\thispagestyle{empty}

{\hfill \bf hep-th/9704068}

\vskip 2.0cm

\begin{center}
{\Large \bf
Constrained twistors principle \\
of the string theory}

\vskip 1.5cm

{\bf A. A. Kapustnikov \footnote{E-mail: alexandr@ff.dsu.dp.ua}
\quad and \quad
S. A. Ulanov \footnote{E-mail: theorph@ff.dsu.dp.ua}}
\vskip 0.5cm
{\it Department of Physics, Dnepropetrovsk University, \\
     320625 Dnepropetrovsk, Ukraine}

\vskip 2cm
{\bf Abstract}
\end{center}
The new principle of constrained twistor-like variables is proposed
for construction of the Cartan 1-forms on the worldsheet of the
$D=3,~4,~6$ bosonic strings. The corresponding equations of motion
are derived. Among them there are two well-known Liouville equations
for real and complex worldsheet functions $W(\ss,\sb)$. The third one
in which $W(\ss,\sb)$ is replaced by the quaternionic worldsheet
function is unknown and can be thought of as that of the $SU(2)$
nonlinear $\sigma$-model governing the classical dynamics of the
bosonic string in $D=6$.

\vskip 0.2cm

\vskip 1cm
\begin{center}
{\it Submitted to Phys. Lett. B}
\end{center}
\vfill
\setcounter{page}0
\setcounter{footnote}0
\newpage

\section{Introduction}

Let us remind that in the geometrical approach to string theories the
classical dynamics of $D=3$ and $D=4$ relativistic strings is
described by the real and complex Liouville equations for the
corresponding worldsheet function $W(\ss,\sb)$ \cite{1}. These
equations are achieved by solving for the Virasoro gauge constraints
in terms of independent physical string variables restricted only by
the embedding conditions. The latter are imposed to insure that the
string worldsheet would be a minimal surface in a target space-time.
It is worth mentioning that the embedding procedure can be regarded
as a further restriction of the string coordinates which involves the
second order worldsheet derivatives. It has been observed by
Barbashov, Nesterenko and Chervyakov ( BNC in what follows ) \cite{2}
that only when these constraints are taken into account together with
the Virasoro ones, the number of the string coordinates in a
$D$-dimensional space-time are reduced to the $2(D-2)$ independent
variables. Just in that case the system of the Maurer-Cartan
equations which provides the embedding of the worldsheet surface into
the target space-time is reduced to the Liouville equation in the
dimensions $3$ and $4$.

In this letter we would like to note that there are nice solutions
for both of the Virasoro and BNC constraints in terms of the
twistor-like variables subjected to the first order derivatives
conditions. It will be shown that these twistor constraints are a new
basic ingredients of any geometrical approach to a string theory,
which at least in the cases of lower non-critical dimensions allows
one to gain a deeper insight into the structure of strings.

The paper is organized as follows.

In Section 2 for the convenience of the reader we deal with the
first-hand view of the alternative formulation of already known
examples for the $D=3$ and $4$ strings. We show how with the help of
constrained twistors principle one can get all the relevant
geometrical objects of the corresponding worldsheet omitting the
tedious technical details of calculations inherent to the traditional
approaches  \cite{3}.

The generalization of these results to the case of six space-time
dimensions is developed in Section 3. Here the new equation of motion
for the $D=6$ strings is derived in terms of the quaternionic
worldsheet function $W(\ss,\sb)$. The latter turns out to be
composed of the constrained twistor-like variables which are
represented by the spinors of the group $SU(4) = SL(2,H) \sim
SO(1,5)$. It is amusing that the l.h.s. of the new equation strongly
resembles the ordinary equation of the nonlinear $\sigma$-model for
the principal field $G=\exp(-W_0-i \vec W \vec \sigma)$ taking its
values on the $SU(2)$ group.

\section{Constrained twistor-like variables in $D=3,~4$}

It is well-known that in $D=3$ space-time the string coordinates
$x^m(\ss,\sb),~m=0,1,2$ are restricted by the following equation of
motion and covariant constraints \cite{1,2}:
  \begin{equation}\label{1}
  \partial_{\ss} \partial_{\sb} x^m = 0
  \end{equation}
-- the equation of motion,
  \begin{equation}\label{2}
  \partial_{\ss} x^m \partial_{\ss} x_m =
  \partial_{\sb} x^m \partial_{\sb} x_m = 0
  \end{equation}
-- the Virasoro conditions,
  \begin{equation}\label{3}
  \begin{array}{c}
  \partial^2_{\ss} x^m \partial^2_{\ss} x_m = - 4 \qq^2,\\
  \partial^2_{\sb} x^m \partial^2_{\sb} x_m = - 4 \qb^2
  \end{array}
  \end{equation}
-- the BNC constraints, where $\qq$ and $\qb$ are the real parameters
of dimension $+1$. It is easy to see from (\ref{1}-\ref{3}) that
there exists the set of commuting $SL(2,R)$ spinors
$v^{\mu}(\ss),~u^{\mu}(\sb),~\mu=1,2$, which allows one to resolve
these constraints in form
  \begin{eqnarray}\label{4}
  && \partial_{\ss} x^{\mu\nu} = v^{\mu} v^{\nu}, \qquad
     \partial_{\sb} x^{\mu\nu} = u^{\mu} u^{\nu}, \nonumber \\
  && v_{\mu} \partial_{\ss} v_{\nu} - v_{\nu} \partial_{\ss} v_{\mu} =
     \qq \e_{\mu\nu}, \\
  && u_{\mu} \partial_{\sb} u_{\nu} - u_{\nu} \partial_{\sb} u_{\mu} =
     \qb \e_{\mu\nu}, \qquad \e^{12} = 1. \nonumber
  \end{eqnarray}
The conventions in (\ref{4}) are as follows
  \begin{equation}\label{5}
   x^{\mu\nu} = \frac{1}{2} x^m \gamma^{\mu\nu}_m
  \end{equation}
with the symmetric $SL(2,R)$ matrices
  \begin{equation}\label{6}
  \gamma^{\mu\nu}_m =
  \left( \begin{array}{cc} ~1 & ~0 \\ ~0 & ~1 \end{array} \right),~
  \left( \begin{array}{cc} ~0 & -1 \\ -1 & ~0 \end{array} \right),~
  \left( \begin{array}{cc} ~1 & ~0 \\ ~0 & -1 \end{array} \right).
  \end{equation}
Now it is not difficult to verify that the worldsheet function
$W(\ss,\sb)$ which is defined with the relation
  \begin{equation}\label{7}
  u_{\mu}(\sb) v_{\nu}(\ss) - u_{\nu}(\sb) v_{\mu}(\ss) =
  \e_{\mu\nu}\exp[-W(\ss,\sb)]
  \end{equation}
satisfies the Liouville equation
  \begin{equation}\label{8}
  \partial_{\ss} \partial_{\sb} W = \qq \| \qb \| \exp(2W).
  \end{equation}
Indeed, let us introduce the set of the harmonic variables
$v^{\pm}_{\mu}(\ss,\sb)$ instead of chiral ones:
  \begin{eqnarray}\label{9}
  && v^+_{\mu}(\ss,\sb) = \exp(\frac{1}{2} W) u_{\mu}(\sb), \nonumber \\
  && v^-_{\mu}(\ss,\sb) = \exp(\frac{1}{2} W) v_{\mu}(\ss), \\
  && v^+_{\mu} v^-_{\nu} - v^+_{\nu} v^-_{\mu} =
     \e_{\mu\nu}. \nonumber
  \end{eqnarray}
Then directly from the constraints (\ref{4}) we get the following
quantities
  \begin{eqnarray}\label{10}
  \Omega^{--} &\equiv& v^{-\mu}dv^-_{\mu}=\qq \| d\ss \| \exp(W),\nonumber \\
  \Omega^{++} &\equiv& v^{+\mu}dv^+_{\mu}=\qb \| d\sb \| \exp(W),\\
  \Omega^{~0~}&\equiv& v^{+\mu}dv^-_{\mu}= -\frac{1}{2}
    (d\ss \| \partial_{\ss} - d\sb \| \partial_{\sb})W,\nonumber
  \end{eqnarray}
which are easily recognized as the differential one-forms that
appeared in the framework of a standard geometrical approach to a
string theory \cite{3,4,5}. Thus, all the results of this approach
can be recast in the harmonic representation (\ref{4}),(\ref{9}). In
particular, to recover the Liouville equation (\ref{8}) it is
sufficient to notice that the forms $\Omega^{\pm\pm,~0}$ as it stands
in (\ref{10}) satisfy the $SL(2,R)$ Maurer-Cartan equations
  \begin{equation}\label{11}
  d\Omega^{\pm\pm} \pm 2 \| \Omega^0 \Omega^{\pm\pm} = 0,
  \end{equation}
  \begin{equation}\label{12}
  d\Omega^{~0~} + \Omega^{++} \Omega^{--} = 0.
  \end{equation}
The first two equations (\ref{11}) are satisfied identically when the
r.h.s. of equations (\ref{10}) are taken into account while the third
one (\ref{12}) is equivalent to (\ref{8}).

Thus we have constructed a new version of the geometrical approach
to a string theory in $D=3$. The guiding principle of this
formulation was the twistor equation (\ref{4}) where the twistor-like
variables $v^{\mu}(\ss),~u^{\mu}(\sb)$ and its first derivatives are
combined as the components of the relativistic basis. We have right
to anticipate that this formulation will be useful in the cases of
higher dimensions too.

To be more precisely let us consider the string in $D=4$ in brief
outline. In this case as it is well-known the Lorentz group is
isomorphic to the $SL(2,C)$.  Accordingly, the constraints (\ref{4})
can be rewritten in terms of the $SL(2,C)$ Van der Warden spinors
$v^{\mu}(\ss),~u^{\mu}(\sb)$ and their conjugated
$\vb^{\mud}(\ss),~\ub^{\mud}(\sb)$:
  \begin{eqnarray}\label{13}
  \partial_{\ss} x^{\mu\mud} = v^{\mu}(\ss) \vb^{\mud}(\ss), &&
     \partial_{\sb} x^{\mu\mud} = u^{\mu}(\sb) \ub^{\mud}(\sb), \nonumber \\
  v_{\mu} \partial_{\ss} v_{\nu} - v_{\nu} \partial_{\ss} v_{\mu} =
     \qq \| \e_{\mu\nu}, &&
     u_{\mu} \partial_{\sb} u_{\nu} - u_{\nu} \partial_{\sb} u_{\mu} =
     \qb \| \e_{\mu\nu}, \\
  x^{\mu\mud} = \frac{1}{2} x^m \sigma^{\mu\mud}_m, &&
     m=0,1,2,3. \nonumber
  \end{eqnarray}
The only difference of equation (\ref{13}) in comparison with
(\ref{4}), (\ref{5}) and (\ref{6}) is that the parameters $\qq,\qb$
become complex and the matrices (\ref{6}) are replaced by the
relativistic $SL(2,C)$ matrices $\sigma^{\mu\mud}_m$.

Their relation to all the other equations (\ref{7}) - (\ref{12}) of
our formulation is just the same as in the case of $D=3$ space-time.
Moreover, the Liouville equation (\ref{8}) is still unchanged too but
with the complex worldsheet function $W(\ss,\sb)$ defined by the
expression formally coinciding with (\ref{7}).

Notice, that as in the case of $D=3$ considered previously, the
chiral twistor-like variables $v_{\mu}(\ss),~u_{\mu}(\sb)$ are
restricted by the equation (\ref{13}) involving its first
derivatives. Once again this equation turns out to be decisive in
reducing the dynamics of string to the nonlinear Liouville equation.

In the next Section similar equation will be used to extend this
approach to more complicated case of $D=6$ string.\footnote{
The case of $D=5$ string is excluded because it yields the known
system of equations which has been discussed earlier (see e.g.
\cite{1}). }

\section{Bosonic string in $D=6$ as a nonlinear $\sigma$-model}

We begin this section with some preliminary remarks concerning our
notations and conventions.

In the case we are dealing with the string coordinates which are
represented by the antisymmetric matrix
  \begin{equation}\label{14}
  x^{\mu\nu} = - x^{\nu\mu} = x^m \gamma^{\mu\nu}_m
  \end{equation}
with the $\gamma$-matrices satisfying the conventions \cite{6,7}
  \begin{equation}\label{15}
  \begin{array}{l}
  \gamma^m_{\mu\nu} \gamma^{n\mu\nu} = - 4 \eta^{mn}, \qquad
     \eta = diag(+-----), \\
  \gamma^m_{\mu\nu} \gamma_{m\rho\sigma}=-2\e_{\mu\nu\rho\sigma}.
  \end{array}
  \end{equation}
Respectively, the underlying twistor-like variables are introduced as
the four-component Weyl spinors $v^{\mu}_i(\ss),~u^{\mu}_i(\sb)$ with
$SU(4) = SL(2,H) \sim SO(1,5)$ Lorentz index $\mu=1,2,3,4$ and an
extra $SU(2)$ doublet index $i=1,2$. Note that because of the
pseudoreality condition
  \begin{equation}\label{16}
  \overline{v^{\mu}_i} = C^{\mud}_{~\nu} \e^{ij} v^{\nu}_j,
  \end{equation}
where $C^{\mud}_{~\nu}$ is the charge conjugation matrix, the spinor
$v^{\mu}_i$ has primarily eight real components. But only half of them
may be considered as truly independent.\footnote{ See also discussion
in \cite{7}.} The rest ones can be eliminated due to the
orthogonality condition
  \begin{equation}\label{17}
  T^{~j}_i \equiv v^{\mu}_i v^j_{\mu} = 0.
  \end{equation}
To prove the last statement let us consider the $SU(2)$ invariant
vector
  \begin{equation}\label{18}
  v^{\mu\nu} = \e^{ij} v^{\mu}_i v^{\nu}_j.
  \end{equation}
This vector is light-like because from the rules of the raising and
lowering the pair of antisymmetric $SU(4)$ indices \cite{6}
  \begin{equation}\label{19}
  v^{\mu\nu} = \frac{1}{2} \e^{\mu\nu\rho\sigma} v_{\rho\sigma}, \qquad
  v_{\mu\nu} = \frac{1}{2} \e_{\mu\nu\rho\sigma} v^{\rho\sigma},
  \end{equation}
the vector (\ref{18}) is constrained to be
  \begin{equation}\label{20}
  v^{\mu\nu} v_{\mu\nu} = - v_{\rho\sigma} v^{\rho\sigma} = 0.
  \end{equation}
This restricts the matrix (\ref{17}) to be singular one because in
accordance with (\ref{18}) the equation (\ref{20}) reads as follows
  \begin{equation}\label{21}
  v^{\mu\nu} v_{\mu\nu} = \e^{ij} \e_{kl} T^{~k}_i T^{~l}_j =
    2 \| det \| T = 0.
  \end{equation}
Thus we conclude that the rows and the columns of the matrix $T$ are
not independent, say
  \begin{equation}\label{22}
  v^{\mu}_1 v_{\mu i} = \alpha v^{\mu}_2 v_{\mu i}
  \end{equation}
for $\alpha$ being constant. On the other hand, for any spinor $v_{\mu
k}$ the identity
  \begin{equation}\label{23}
  v^{\mu\nu} v_{\nu k} = \frac{1}{2} \e^{\mu\nu\rho\sigma} \e^{ij}
    v_{\rho i} v_{\sigma j} v_{\nu k} = 0
  \end{equation}
can be derived since the $SU(2)$ indices $i,j,k$ run over two values
only. Substituting the condition (\ref{22}) in (\ref{23}) we can
rewrite equation (\ref{22}) in the following form
  \begin{equation}\label{24}
  (v^{\mu}_1 - \alpha v^{\mu}_2) v^{\nu}_2 v_{\nu k} = 0.
  \end{equation}
It is rather evident now that equation (\ref{22}) has only one
acceptable solution (\ref{17}). The alternative one $v^{\mu}_1=\alpha
v^{\mu}_2$ is not reasonable since in virtue of this solution the
vector $v^{\mu\nu}$ turns out to be equal to zero identically.

Now we are ready to proceed to the construction of the twistor-like
solutions of the equations (\ref{1})-(\ref{3}) for the $D=6$ string.
Here we shall use general conventions on the twistor basis in $D=6$
which turns out to be very useful, especially for getting the basis
appropriated for the spinor harmonic variables.

We can say that the pair of spinors $v^{\mu}_i,~w^{\mu}_j$ forms
the basis in six dimensions if they satisfy the following
orthonormality conditions
  \begin{equation}\label{25}
  v_{\mu i} w^{\mu}_j = v^{\mu}_i w_{\mu j} = \e_{ij}, \qquad
  \e^{ij} ( v^{\mu}_i w_{\nu j} + v_{\nu i} w^{\mu}_j ) =
    - \delta^{\mu}_{\nu}.
  \end{equation}

Apparently, this notion is crucial for our further investigation.
Indeed, let us consider the twistor-like solutions to the Virasoro
constraints (\ref{2}) in $D=6$
  \begin{equation}\label{26}
  \partial_{\ss} x^{\mu\nu} = v^{\mu\nu}, \qquad
  \partial_{\sb} x^{\mu\nu} = u^{\mu\nu} =
    \e^{ij} u^{\mu}_i(\sb) u^{\nu}_j(\sb),
  \end{equation}
with $x^{\mu\nu}(\ss,\sb)$ and $v^{\mu\nu}(\ss)$ defined in equations
(\ref{14}) and (\ref{18}) respectively. It can be shown that in these
notations the BNC conditions (\ref{3}) for $D=6$ string may be
represented in form
  \begin{equation}\label{27}
  \e_{ij} \e_{kl} \e ^{\mu\nu\rho\sigma} v^i_{\mu} \| v^j_{\nu} \|
    \partial_{\ss} v^k_{\rho} \| \partial_{\ss} v^l_{\sigma} =
    - 16 \| \qq^2
  \end{equation}
(the same is true for the spinor $u^{\mu}_i(\sb)$). At the same time,
taking account of (\ref{25}) one can get the equation
  \begin{equation}\label{28}
  \e_{ij} \e_{kl} \e ^{\mu\nu\rho\sigma} v^i_{\mu} \| v^j_{\nu} \|
    w^k_{\rho} \| w^l_{\sigma} = 4.
  \end{equation}
We see that both of the equations (\ref{27}) and (\ref{28}) have
the same structure. It suggests the following twistor-like solution
to the BNC constraints in six dimensions
  \begin{equation}\label{29}
  \partial_{\ss} v^{\mu}_i = 2 \| i \| \qq \| w^{\mu}_i(\ss), \qquad
  \partial_{\sb} u^{\mu}_i = 2 \| i \| \qb \| \omega^{\mu}_i(\sb).
  \end{equation}
Thus we obtain the set of equations
  \begin{eqnarray}\label{30}
  & v_{\mu i} \partial_{\ss} v^{\mu}_j = 2 \| i \| \qq \| \e_{ij}, \qquad
  u_{\mu i} \partial_{\sb} u^{\mu}_j = 2 \| i \| \qb \| \e_{ij},&\nonumber \\
  & \e^{ij} ( v^{\mu}_i \| \partial_{\ss} v_{\nu j} +
             v_{\nu i} \| \partial_{\ss} v^{\mu}_j ) =
    - 2 \| i \| \qq \| \delta^{\mu}_{\nu}, & \\
  & \e^{ij} ( u^{\mu}_i \| \partial_{\sb} u_{\nu j} +
             u_{\nu i} \| \partial_{\sb} u^{\mu}_j ) =
    - 2 \| i \| \qb \| \delta^{\mu}_{\nu}, & \nonumber
  \end{eqnarray}
which provides the presence of the twistor-like basis in six
dimensions similar to (\ref{4}) and (\ref{13}).

The prescriptions to be used in Section 2 can be straightforwardly
generalized now to describe an embedding of the strings worldsheet
into the $D=6$ flat space-time. Namely, the equation (\ref{7})
acquires the form
  \begin{equation}\label{31}
  \begin{array}{c}
  u^{\mu}_i v^j_{\mu} = G^{~j}_i \equiv [ \exp(-W) ]^{~j}_i,\\
  {G^{-1}}^{~j}_i (u^{\mu}_j v^i_{\nu} + u_{\nu j} v^{\mu i}) =
    \delta^{\mu}_{\nu}.
  \end{array}
  \end{equation}

Note that the $SU(2)$ matrix function $W^{~j}_i(\ss,\sb)$ that appears
in (\ref{31}) is not arbitrary. From the pseudoreality conditions
(\ref{16}) it follows that
  \begin{equation}\label{32}
  \overline{G^{~j}_i} = - \e^{ik} \e_{jl} G^{~l}_k.
  \end{equation}
Equation (\ref{32}) restricts the matrix $W^{~j}_i$ to take its
values in the $SU(2)$ group
  \begin{equation}\label{33}
  W^{~j}_i = W_0 \delta^j_i + i \| \vec W \vec \sigma^{~j}_i
  \end{equation}
with $W_0$ and $\vec W$ being real. So, the new feature of the $D=6$
string contrarily to above mentioned examples is that the
corresponding worldsheet function $W(\ss,\sb)$ is represented by the
quaternionic field (\ref{33}).

To get the associated equation of motion, the following redefinition
of the twistor variables will be done
  \begin{equation}\label{34}
  v^{\mu +}_i = {G^{-1/2}}^{~j}_i u^{\mu}_j, \qquad
  v^{\mu -}_i = v^{\mu j} {G^{-1/2}}_{ji}.
  \end{equation}
Then equation (\ref{31}) ensures the ordinary orthonormality
properties of the harmonic variables (\ref{34}), namely
  \begin{eqnarray}\label{35}
  & v^{\mu +}_i v^{- j}_{\mu} = \delta^j_i, & \nonumber \\
  & v^{\mu \pm}_i v^{\pm j}_{\mu} = 0, & \\
  & \e^{ij} ( v^{\mu +}_i v^{-}_{\nu j} +
    v^{+}_{\nu i} v^{\mu -}_j ) = \delta^{\mu}_{\nu}. & \nonumber
  \end{eqnarray}
The corresponding Cartan forms derived from (\ref{30}), (\ref{34}),
(\ref{35}) are
  \begin{eqnarray}\label{36}
  \Omega^{--} &\equiv& v^{\mu-} dv^{-}_{\mu} =
    - 2 \| i \| \qq \| d\ss \| \tilde G \| G^{-1}, \nonumber \\
  \Omega^{++} &\equiv& v^{\mu+} dv^{+}_{\mu} =
    - 2 \| i \| \qb \| d\sb \| G^{-1} \tilde G, \\
  \Omega^{~0~} &\equiv& v^{\mu +} dv^{-}_{\mu} =
    G^{-1/2} d\ss \| \partial_{\ss} G^{ 1/2} +
    G^{ 1/2} d\sb \| \partial_{\sb} G^{-1/2}, \nonumber
  \end{eqnarray}
where $\tilde G = \exp(-i \vec W \vec \sigma)$ is the $SU(2)$-valued
field. The 1-forms (\ref{36}) can be viewed as a nonabelian
generalization of the connection forms (\ref{10}) from our earlier
discussion. In accordance with their definition $\Omega^{0,\pm\pm}$
satisfy the Gauss equation
  \begin{equation}\label{37}
  d\Omega^{0} - \Omega^{++} \Omega^{--} + \Omega^{0} \Omega^{0} = 0
  \end{equation}
which connects the fields in the r.h.s. of equation (\ref{36}) with
the following equation of motion
  \begin{equation}\label{38}
  \partial_{\sb} ( \partial_{\ss} G^{-1/2} G^{ 1/2} ) -
  \partial_{\ss} ( \partial_{\sb} G^{ 1/2} G^{-1/2} ) +
  [ G^{-1/2} \partial_{\ss} G^{1/2}, G^{1/2} \partial_{\sb} G^{-1/2} ] =
    4 \| \qq \| \qb \| G^{-1} \tilde G^{\| 2} G^{-1}.
  \end{equation}
Thus, the dynamics of the $D=6$ string is described by the following
pair of the equations of motion
  \begin{equation}\label{39}
  \partial_{\ss} \partial_{\sb} W_0 = 4 \| \qq \| \qb \| e^{2W_0},
  \end{equation}
  \begin{equation}\label{40}
  \partial_{\sb} ( \partial_{\ss} \tilde G^{-1/2} \tilde G^{ 1/2} ) -
  \partial_{\ss} ( \partial_{\sb} \tilde G^{ 1/2} \tilde G^{-1/2} ) +
  [ \tilde G^{-1/2} \partial_{\ss} \tilde G^{1/2},
    \tilde G^{1/2}  \partial_{\sb} \tilde G^{-1/2} ] = 0.
  \end{equation}
Note that despite the difference in sign in the l.h.s. of the
equations (\ref{37}) and (\ref{12}) the sign in r.h.s. of the
equation (\ref{39}) appears to be compatible with that of the
Liouville equation (\ref{8}). This is the result of the normalization
condition (\ref{30}) which in comparison with (\ref{4}), (\ref{13})
is supplemented by the multiplier $i$.

\section{Conclusion}

Thus we have demonstrated here that the constrained twistors
principle enjoys a wide-spread support in the framework of the string
theory. As opposed to the known geometrical approaches it provides us
with the twistor-like variables compatible with the Virasoro and BNC
constraints. These variables have the additional benefit allowing
simultaneous treatment of the lower-dimensional strings after
replacing the real $SL(2,R)$ spinors by the $SL(2,C)$ complex or
$SL(2,H)$ pseudoreal ones, one recovers the connection forms
associated with the zero curvature representations of the three-,
four- or six-dimensional strings. On the other hand, parametrizing
these forms by the real, complex or quaternionic worldsheet function
$W(\ss,\sb)$ we obtain the new equation of motion for $D=6$ string
besides the known Liouville equations. As far as we know, neither
the Cartan forms (\ref{36}) nor the equation (\ref{38}) describing
the dynamics of the $D=6$ string in such a minimal form have been
obtained yet. Note that the $SO(D-2)$ nonlinear $\sigma$-model
proposed in \cite{5} could not be thought of as a minimal one because
even in the case of $D=6$ it keeps a number of the redundant fields
which could not be eliminated with the help of any reasonable
dynamical principle.

In this article we have dwelled upon the bosonic strings only. All
the important peculiarities of our approach are assumed to be
applicable in the superstring theory. It can be proved, for instance,
that to get a new version of the supersymmetric generalization of the
Liouville equation given recently by Bandos, Sorokin and Volkov
\cite{5} it is sufficient to replace the constrained $SL(2,R)$
twistors of Section 2 with its supersymmetric (anti)chiral
superfields counterparts. We hope also that this direction of study
may turn out to be instructive for a better understanding of the
structure of the WZWN models related to the superstring theory
\cite{8,9}.

\vskip 0.5cm
{\bf Acknowledgments}

The authors would like to thank Dr. A. Pashnev for careful reading
of the manuscript and a numerous of the critical remarks.

This work was supported in part by ISSEP grant APU062045 and INTAS
grant 94-2317.

\newpage


\def\VYP #1 #2 #3{{\bf #1} (#2) #3}
\def\AVYP #1 #2 #3 #4{{\bf #1} \VYP {#2} {#3} {#4}}

\def\Jtit #1{{\sl #1}}
\def\Journ #1 #2 #3 #4{\Jtit {#1}~ \VYP {#2} {#3} {#4} }
\def\JournA #1 #2 #3 #4 #5{\Jtit {#1}~ \AVYP {#2} {#3} {#4} {#5}}
\def\Aut #1{{\bf #1}}
\def\Jitem #1 #2 #3 #4 #5{\Aut {#1}, \Journ {#2} {#3} {#4} {#5}}
\def\JitemA #1 #2 #3 #4 #5 #6{\Aut {#1}, \JournA {#2} {#3} {#4} {#5} {#6}}

\def\CMPh {Commun. Math. Phys.}
\def\JPh  {J. Phys.}
\def\LMPh {Lett. Math. Phys.}
\def\NPh  {Nucl. Phys.}
\def\PhL  {Phys. Lett.}
\def\PhR  {Phys. Rev.}
\def\PhRL {Phys. Rev. Lett.}

\def\Book #1 #2 #3 #4 {\Aut {#1} {\sl #2}~~#3,~#4}

\end{document}